\documentclass[fleqn,usenatbib]{mnras}
\usepackage{newtxtext,newtxmath} 
\usepackage[pdftex]{graphicx}
\usepackage[pdftex]{color}

\usepackage[T1]{fontenc} 
\DeclareRobustCommand{\VAN}[3]{#2}
\let\VANthebibliography\thebibliography
\def\thebibliography{\DeclareRobustCommand{\VAN}[3]{##3}\VANthebibliography}

\usepackage{amsmath}	% Advanced maths commands
\usepackage{url}
 \def\deg{^\circ} 
 \def\deg{^\circ}
 \def\/{\over}\def\kms{km s$^{-1}$}

 \def\be{\begin{equation}} 
 \def\ee{\end{equation}}
 \def\kms{km s$^{-1}$}    
  \def\Htwo{H$_2$ }  
\def\({\left(} \def\){\right)} \def\[{\left[} \def\]{\right]}

 \def\Tex{T_{\rm ex}}

\def\NHtwo{N_{\rm H_2}} \def\Ncoth{N_{\rm ^{13}CO}}
\def\Ycoth{Y_{\rm ^{13}CO}}
\def\htwo{H$_2$} 
\def\uxco{[\Htwo cm$^{-2}$(K \kms)$^{-1}$\ } 
 
\def\Tbcotw{T_{\rm B}({\rm ^{12}CO})} 
\def\Tbcoth{T_{\rm B}({\rm ^{13}CO})} 

\def\scx{\rm SCD_{\rm 12X}} 
\def\scl{\rm SCD_{\rm 13L} }

\def\Xgeneral{X_{\rm CO}}

\def\revone{}
\title[CO-to-H$_2$ Conversion Factor II]{The CO-to-H$_2$ Conversion Factor of Galactic Giant Molecular Clouds using CO isotopologues: High-resolution $\Xgeneral$ maps }
 \author[M. Kohno and Y. Sofue]{
Mikito Kohno$^{1,2}$\thanks{E-mail: mikito.kohno@gmail.com, kohno@nagoya-p.jp}, and
Yoshiaki Sofue$^{3}$
\\
$^{1}$Astronomy Section, Nagoya City Science Museum, 2-17-1 Sakae, Naka-ku, Nagoya, Aichi 460-0008, Japan\\
$^{2}$Department of Physics, Graduate School of Science, Nagoya University, Furo-cho, Chikusa-ku, Nagoya, Aichi 464-8602, Japan\\
$^{3}$Institute of Astronomy, The University of Tokyo, Mitaka, Tokyo 181-0015, Japan}

\date{Accepted 2023 November 22. Received 2023 November 1; in original form 2023 May 9} 
\pubyear{2024}
\begin{document} 

\maketitle

\begin{abstract}  
{%\sof{\underline{適宜magentaで加筆・コメントしました。}}\\
%\sof{\underline{png図のサイズが大きいのでコンパイルに時間がかかりすぎます。}}\\
%\sof{\underline{pdf/jpgにしてみたら図の大きさは変わりませんが少し軽くなりました。}}\\
%\sof{\underline{図３背景をも少し暗くしたらどうでしょう。}}\\
%\sof{\underline{Eq. 16, 17当たりの議論、自身ありません。}}\\
%\sof{\underline{逆センスのような気がしてきました。整理してくれませんか？}}\\
We investigated the correlation between intensities of the $^{12}$CO and $^{13}$CO ($J=1$-0) lines toward the Galactic giant molecular clouds (GMCs) W51A, W33, N35-N36 complex, W49A, M17SW, G12.02-00.03, W43, and M16 using the FUGIN (FOREST Unbiased Galactic plane Imaging survey with the Nobeyama 45-m telescope) CO line data. 
All the GMCs {show} intensity saturation in the $^{12}$CO line when the brightness temperature of $^{13}$CO is higher than a threshold temperature of about $\sim 5$ K.
We obtained high-resolution ($\sim 20\arcsec$) distribution maps of the $\Xgeneral$ factor ({$X_{\rm CO, {iso}}$}) in individual GMCs using correlation diagrams of the CO isotopologues.
It {is} shown that {$X_{\rm CO, {iso}}$} is variable in each GMC within the range of $X_{\rm CO, {iso}} \sim (0.9 {\rm -} 5) \times 10^{20}$ cm$^{-2}$ (K\ km\ s$^{-1})^{-1}$.
Despite the variability in the GMCs, the {average} value among the GMCs {is} found to be nearly constant at $X_{\rm CO, {iso}} = {(2.17 \pm 0.27) \times 10^{20}}$ cm$^{-2}$ (K\ km\ s$^{-1})^{-1}$, which is consistent with that from previous studies in the Milky Way.
}
%FUGIN CO survey dataを用いて、銀河面の巨大分子雲 W51, W33, N35-N36 complex, W49, M17SW, M17SWex について、$^{12}$CO, $^{13}$CO の correlation 関係を調べ、$\Xgeneral$ factor を求めた。
%$^{12}$CO は、$^{13}$COのbrightness temperature が高くなるについて、saturation する
%傾向が共通して見られた。$^{12}$COがdiffuse に広がったN35-N36 complex, M17 Swex では、saturation の影響が大きく、$\Xgeneral\sim 3 \times 10^{20}$とMilky Way の平均値より大きい値を取ることがわかった。各領域ごとのXCO ばらつきの原因は星間ガスの構造の違いに起因すると考えられる。
%また、得られたX-factor の全体の平均値は$\Xgeneral = 2.38 \pm 0.47$ [cm$^{-2}$ (K\ km\ s$^{-1})^{-1}$]であり、これまでの研究とエラーの範囲内で一致することがわかった。
\end{abstract}


\section{Introduction} 
{Giant} molecular clouds (GMC) mainly consist of hydrogen molecules (H$_2$) and are the sites of star formation in galaxies \citep{2007prpl.conf...81B,2010ARA&A..48..547F,2014prpl.conf....3D,2022arXiv220309570C}. 
The hydrogen molecule is difficult to observe directly because it {has no} an electric dipole moment for its homonuclear diatomic nature. 
Therefore, we often observe carbon monoxide (CO) with rotational transitions at millimeter wavelengths \citep{2015ARA&A..53..583H} and convert the intensity to the H$_2$ column density using the CO-to-\htwo\ conversion factor (hereafter $X_{\rm CO}$), which is given by
\begin{equation}
X_{\rm CO} = N_{\rm H_2}/W_{\rm ^{12}CO} \ {\rm [cm^{-2}\ (K\ km\ s^{-1})^{-1}]},
\label{XCO}
\end{equation}
where $W_{\rm ^{12}CO}$ and $N_{\rm H_2}$ are the $^{12}$CO $J=$1-0 integrated intensity and column density of hydrogen {molecules}, respectively \citep{2013ARA&A..51..207B}.

The conversion factor has been obtained in various {methods by comparing} CO luminosity with
virial mass (e.g.,\citealp{1987ApJ...319..730S,1987ApJS...63..821S}), 
visual or infrared extinction (e.g.,\citealp{1982ApJ...262..590F,2006A&A...454..781L,2008ApJ...679..481P,2018MNRAS.474.4672L}), 
X-ray absorption (e.g.,\citealp{Sofue+2016}), {gamma-ray brightness}(e.g.,\citealp{1988A&A...207....1S,Abdo+2010,Planck+2015,Hayashi+2019b}), 
and dust emission (e.g.,\citealp{2011A&A...536A..19P,Leroy+2011,2013ApJ...777....5S,2014ApJ...796...59F,Okamoto+2017,Hayashi+2019a,2023PASJ..tmp...51Y}).
The values have been discussed in the Milky Way and Local Group galaxies, and {their dependence} on metallicity has been suggested (e.g.,\citealp{1995ApJ...448L..97W,arimoto+1996,1996ApJ...462..215S,2001ApJ...547..792D,2001PASJ...53L..45M,2003ApJ...599..258R,2010ARA&A..48..547F,2010A&A...518A..45L,Leroy+2011,2014ApJ...784...80L,2016ApJ...826..193L,2017ApJ...844...98M,2021ApJS..256....3P,2023ApJ...949...63O}).
The often used value currently in the Milky Way is $X_{\rm CO} = 2.0 \times 10^{20}\ {\rm cm^{-2}\ (K\ km\ s^{-1})^{-1}}$, having {an} uncertainty of $\pm 30\%$ \citep{2013ARA&A..51..207B}.

Recently, \cite{2020MNRAS.497.1851S}(hereafter paper I) proposed
a new method to estimate the molecular cloud mass considering the variability of the $\Xgeneral$ factor from the correlation between $^{12}$CO and $^{13}$CO. 
Paper I analyzed only two Galactic GMCs of M16 \citep{2020MNRAS.492.5966S,2021PASJ...73S.285N} and W43\citep{2019PASJ...71S...1S,2021Galax...9...13S,2021PASJ...73S.129K}, {while} the spatial distributions of $\Xgeneral$ inside the GMCs have not {yet been discussed}.

In this paper, we extend the analysis of Paper I, {aiming to reveal} the variability of $X_{\rm CO}$ within the molecular cloud of scales ($\sim 10$-50 pc) at sub-pc resolutions ($\sim 20''$). 
The target GMCs are W51A
\citep{2019PASJ..tmp...46F}, W33 \citep{2018PASJ...70S..50K,2022MNRAS.510.1106M}, W49A \citep{2022PASJ...74..128M}, N35-N36 complex \citep{2018PASJ...70S..51T,2019PASJ...71..104S,2019PASJ...71..121S}, M17SW \citep{2018PASJ...70S..42N,2020PASJ...72...21S,2022MNRAS.509.5809S}, G012.02-00.03\citep{2014ApJ...781..108S}, W43\citep{2019PASJ...71S...1S,2021PASJ...73S.129K}, and M16\citep{2020MNRAS.492.5966S,2021PASJ...73S.285N}. 
These GMCs are also known as massive star-forming regions in the Milky Way.
The basic parameters of these GMCs {are} summarized in Table \ref{tab1}.

This paper is structured as follows: 
section 2 introduces the FUGIN (FOREST Unbiased Galactic plane Imaging survey with the Nobeyama 45-m telescope) data; 
section 3 presents the methods of analysis; 
in section 4, we demonstrate the results; 
in section 5, we discuss the variability of the $X_{\rm CO}$ factor, 
and in section 6, we summarize the results.

%CO-to-\htwo\ conversion factor, $X_{\rm CO}$は、
%分子雲の質量を見積もる方法として、広く使われてきた。
%$^{12}$COのintegrated intensity $W_{\rm ^{12}CO}$とcolumn density $N_{\rm H_2}$を用いて以下の式で表される。 \citep{2013ARA&A..51..207B}
%\begin{equation}
%X_{\rm CO} = {N_{\rm H_2} \over W_{\rm ^{12}CO}} \sim 2.0\times 10^{20}\ {\rm [cm^{-2} (K km\ s^{-1})^{-1}]}
%\label{XCO}
%\end{equation}
%virial mass (e.g.,\citealp{1987ApJ...319..730S})やX-ray (e.g.,\citealp{Sofue+2016}), gamma-ray (e.g.,\citealp{1988A&A...207....1S,Planck+2015,Hayashi+2019b}), dust emission (e.g.,\citealp{2014ApJ...796...59F,Okamoto+2017,Hayashi+2019a})とCO intensityの比較から、Milky Way やLocal group galaxy で、その値が広く議論されてきた。 (e.g.,\citealp{Arimoto+1996,2001PASJ...53L..45M,2010ARA&A..48..547F,Leroy+2011,2017ApJ...844...98M})

%\cite{2020MNRAS.497.1851S}(hereafter paper I)は、$\Xgeneral$とLTE mathodを用いて
%M16 \citep{2020MNRAS.492.5966S,2021PASJ...73S.285N}と
%W43 GMC \citep{2019PASJ...71S...1S, 2021PASJ...73S.129K}
%に対する解析から$^{12}$COのintegrated intensity から真の分子雲の質量を求める新しい手法を提案した。
%しかし、解析は2天体に留まっており、XCO 変動の起源は未解明のままだった。

%本論文では、$\Xgeneral$のvaliability の起源を解明するため、
%paper Iの解析手法を、銀河面のGMCである
%W51 \citep{2019PASJ..tmp...46F}, 
%W33 \citep{2018PASJ...70S..50K}, W49 \citep{2009PASJ...61...39M}, 
%N35-N36 complex \citep{2018PASJ...70S..51T,2019PASJ...71..104S}, 
%M17 \citep{2018PASJ...70S..42N,2020PASJ...72...21S}に拡張して行った。
%各領域のbasic parameters はTable \ref{tab1}にまとめた。\\

\begin{table*} %%%%%%%%%%%%%%%%%%  
\begin{center}
\caption{Properties of the Galactic massive star forming regions}
\begin{tabular}{ccccccccc} 
\hline 
\hline   
Name & $l$ & $b$ & $V_{\rm LSR}$ & $D$ & $R_{\rm G}$  & $T_{\rm rms}$ ($^{13}$CO$J=$1-0)& {$T_{\rm rms}$ ($^{12}$CO$J=$3-2)} & Reference \\
 & [deg] & [deg] & [km s$^{-1}$] & [kpc]  & [kpc] &[K]&[K] &\\
 (1) & (2) & (3) & (4) & (5) & (6) & (7) & (8) & (9) \\
\hline
W51A & $49.5$ & $-0.4$ & 57 & 5.4 & 6.1 & $\sim 0.6$ &{$\sim 0.9$} &[1]\\
W33 & $12.8$ & $-0.2$ & 35 & 2.4 & 5.7  & $\sim 0.6$ & {$\sim 0.4$} &[2]\\
N35-N36 complex & $24.5$ & $0.1$ & 110 & 7.3 & 3.3 &$\sim 0.6$ & {$\sim 0.5$} &[3,4]\\
W49A  & $43.16$ & $0.0$ & 11 & 11& 7.5 & $\sim 0.5$ & {$\sim 0.4$} &[5]\\
M17SW  & $15.0$ & $-0.7$ & 20 & 2.0 & 6.1 & $\sim 0.7$& ---&[6,7,8]\\
G012.02-00.03 & $12.02$ & $-0.03$ & 112 & 9.4 & 2.3  &$\sim 0.7$& {$\sim 0.3$} &[9]\\
W43 & $30.8$ & $0.0$ & 95 & 5.5 &  4.3  &$\sim 0.9$& {$\sim 0.4$} &[10,11]\\
M16 & $17.1$ & $0.7$ & 23 & 2.0 &  6.1  &$\sim 0.7$ & --- &[12,13]\\
\hline    
\end{tabular} \\
\end{center}  
Columns: (1) Name. (2) Galactic longitude (3) Galactic latitude (4) Radial velocity (5) Distance from the solar system (6) Distance from the Galactic center. {The distance to the Galactic centre from the solar system} is assumed to be $R_0=8.0$ kpc obtained by the mean value of the very long baseline interferometry (VLBI) astrometry results \citep{2020PASJ...72...50V,2019ApJ...885..131R}. 
%\sof{(GC distance of the Sun is assumed to be $R_0=8.0$ kpc or 8.2 kpc? }
(7) r.m.s noise {levels} of the $^{13}$CO $J=$1-0 data (8) r.m.s noise {levels} of the $^{12}$CO $J=$3-2 data (9) References [1] \cite{2019PASJ..tmp...46F} [2] \cite{2018PASJ...70S..50K} [3] \cite{2018PASJ...70S..51T} [4] \cite{2019PASJ...71..104S} [5] \cite{2022PASJ...74..128M} [6] \cite{2018PASJ...70S..42N} [7] \cite{2020PASJ...72...21S} [8] \cite{2016ApJ...833..163Y} [9] \cite{2014ApJ...781..108S} [10] \cite{2019PASJ...71S...1S} [11]\cite{2021PASJ...73S.129K} [12] \cite{2020MNRAS.492.5966S}[13] \cite{2021PASJ...73S.285N}
\label{tab1} 
\end{table*} 
%%%%%%%%%%%%%%%%%%%
 
\section{Data} 
%\subsection{The FUGIN project} 
We utilized the $^{12}$CO {and} $^{13}$CO $J=$1-0 line data obtained 
%by FOREST Unbiased Galactic plane Imaging survey 
with the Nobeyama 45 m telescope (FUGIN: \citealp{2017PASJ...69...78U,2019PASJ...71S...2T,2023PASJ...75..279F}). 
{The rest frequencies of $^{12}$CO and $^{13}$CO $J=$1-0 are 115.271 GHz and 110.201 GHz, respectively.}
The front end was the FOur beam REceiver System on the 45-m Telescope (FOREST: \citealp{2016SPIE.9914E..1ZM,2019PASJ...71S..17N}), which is the four-beam, side-band separating (2SB), and dual-polarization superconductor-insulator-superconductor (SIS) receiver. 
The observations were {performed} in the on-the-fly mapping mode \citep{2008PASJ...60..445S}.
The back-end system {was} used an FX-type spectrometer named SAM 45 \citep{Proc..2011,2012PASJ...64...29K}. 
{The half-power beam width (HPBW) of the 45 m telescope is
14\arcsec and 15\arcsec  at 115 GHz and 110 GHz, respectively. The data are gridded to 8.5\arcsec and 0.65 \kms for the spatial and velocity space. The final 3D cube used in this {study} has a voxel size of $(l, b, v) = (8.5$\arcsec$, 8.5$\arcsec, $0.65$ \kms). The effective resolution convolved with a Bessel $\times$ Gaussian function is 20\arcsec and 21\arcsec for $^{12}$CO and $^{13}$CO $J=$1-0, respectively. }
{We subtracted {the} baselines {using} a first-order polynomial function from each spectrum. The baseline ranges {are} adopted from $-200$ \kms to $-50$ \kms and $200$ \kms to $350$ \kms for the first Galactic quadrant with the FUGIN CO survey \citep{2017PASJ...69...78U}.}
{The data {are} calibrated from the antenna temperature ($T_A^*$) to the main-beam temperature ($T_{\rm MB}$) by measuring the main-beam efficiency of 0.43 for $^{12}$CO, 0.45 for $^{13}$CO, and C$^{18}$O using standard calibration sources.}
{Detailed} information on the FUGIN project {is} summarized in the project overview paper by \citet{2017PASJ...69...78U}. 
{Cube} data calibrated to the main-beam temperature 
{are available} at the Japanese Virtual Observatory (JVO).
{We also used the $^{12}$CO ($J=$3-2) High-Resolution Survey (COHRS) archival data of the Galactic Plane obtained by the {James} Clark Maxwell Telescope \citep{2013ApJS..209....8D,2023ApJS..264...16P}. The cube data {are} converted $T_A^*$ to the $T_{\rm MB}$ scale using {a} main-beam efficiency of 0.61 \citep{2023ApJS..264...16P}. The spatial and velocity {resolutions are} 16.6\arcsec and 0.635 \kms, respectively.
Comparing the FUGIN CO survey data, we smoothed them to 20\arcsec with the same spatial resolution.  The $^{12}$CO $J=$3-2 data {are} re-gridded to the same voxel size {as} the FUGIN data using CASA \citep{2022PASP..134k4501C}. {Table \ref{tab1} presents} the root-mean-square noise level of each GMC at the $T_{\rm MB}$ scale.}

%CO data はpaper I と同様にFOREST Unbiased Galactic plane Imaging survey with the Nobeyama 45 m telescope (FUGIN: \citealp{2017PASJ...69...78U,2019PASJ...71S...2T}) projectで取得された$^{12}$CO, $^{13}$CO $J=$1-0 line データを用いた。
%front-endは、野辺山にinstall されたfour-beam, side-band separating (2SB), dual-polarization のSIS receiver であるFOur beam REceiver System on the 45-m Telescope (FOREST: \citealp{2016SPIE.9914E..1ZM,2019PASJ...71S..17N})を使用した。
%観測はon-the-fly mapping mode \citep{2008PASJ...60..445S}で行われた。
%back-end は、digital spectrometer SAM 45 \citep{Proc..2011}を用いた。
%FUGIN project の詳細は、project overview paper \citep{2017PASJ...69...78U}に記載されている。
%なお、解析したcube dataはJapanse Virtual Observartory (JVO) で公開されているものである。

%%%%%%%%%%%%%%%% 
\begin{figure*}
\begin{center}        
\includegraphics[width=15.5cm]{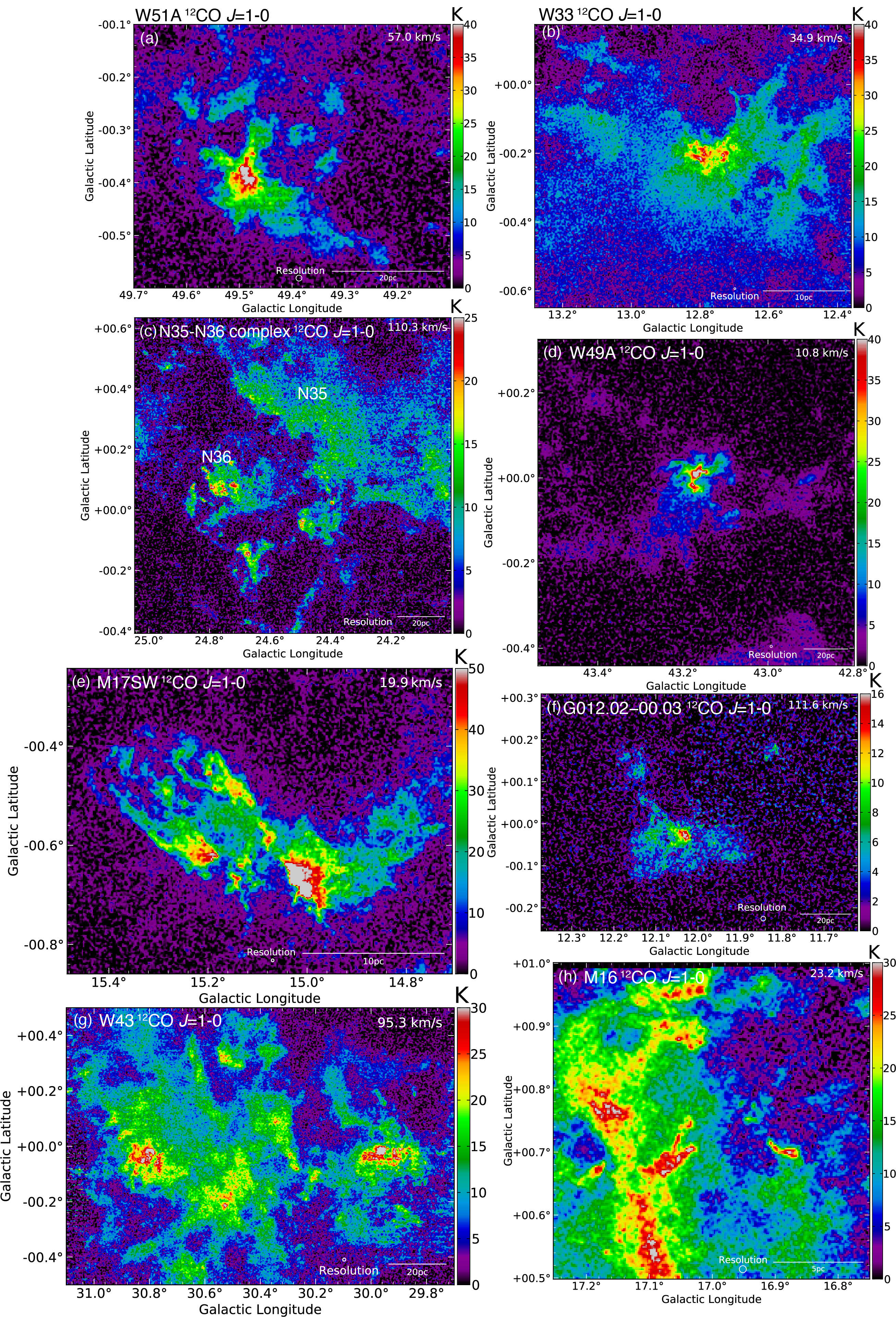}
\end{center}
\caption{The $^{12}$CO peak intensity {maps} of (a) W51A at the radial velocity of 57.0 \kms, (b) W33 at 34.9 \kms, (c) N35-N36 complex at 110.3 \kms, (d) W49A at 10.8 \kms, (e) M17SW at 19.9 \kms, (f) G012.02-00.03 at 111.6 \kms,(g) W43 at 95.3 \kms, and (h) M16 at 23.2 \kms.} 
\label{GMCmap}
\end{figure*}
%%%%%%%%%%%%%%%%%%    

\section{Methods}
We {derived} the H$_2$ column density per velocity channel {using} the local thermal equilibrium (LTE) method {and} the $\Xgeneral$ factor{, as described} in Paper I and \citet{2008ApJ...679..481P}.
The brightness temperature ($T_{\rm B}$) of {the} CO line intensity with the excitation temperature ($T_{\rm ex}$) and optical depth ($\tau$) is given by 
\begin{equation}
T_{\rm B} = T_0
\left(\frac{1}{e^{T_0/\Tex}-1} - \frac{1}{e^{T_0/T_{\rm bg} }-1}\right)
\left(1-e^{-\tau}\right)\ [{\rm K}],
\label{Tb}
\end{equation}
where $T_{\rm bg}=2.725$ K is the temperature of the cosmic microwave background radiation.
$T_0=h \nu/k$ is the Planck temperature with $h$, $\nu$, and $k$ being the Planck constant, rest frequency, and {Boltzmann} constant, respectively. 
If we assume {that} the $^{12}$CO line is optically thick, the excitation temperature is given by
\begin{equation}
\Tex=T_{0}^{115} \bigg/ {\rm ln} \( 1+\frac{T_{0}^{115}}{\Tbcotw_{\rm max}+0.83632}  \)\ [{\rm K}],
\label{Tex}
\end{equation}
where $\Tbcotw_{\rm max}$ and $T_{0}^{115}=5.53194$ correspond to the $^{12}$CO peak intensity and the Planck temperature at the rest frequency of $^{12}$CO $J=$1-0, respectively. We assume that $T_{\rm ex}$ is equal in the $^{12}$CO and $^{13}$CO line emissions, and express the optical depth {as follows:}
\be
\tau(^{13}{\rm CO})=-{\rm ln} 
\(1-\frac{\Tbcoth_{\rm max}/T_{0}^{110} }{(e^{T_{0}^{110}/\Tex}-1)^{-1}- 0.167667} \),
\ee
where $\Tbcoth_{\rm max}$ and $T_{0}^{110}=5.28864$ represent the $^{13}$CO peak intensity and the Planck temperature at the rest frequency of $^{13}$CO $J=$1-0, respectively {\citep{2008ApJ...679..481P}}. 
{{According} to \citet{2009tra..book.....W}, the $^{13}$CO column density is given by}
\be
\Ncoth=3.0\times 10^{14} ~ \frac{\tau}{1-e^{-\tau}}
\frac{1}{ 1-e^{-T_{0}^{110}/\Tex}} ~ I_{\rm ^{13}CO}\ [{\rm cm^{-2}}],
\ee
where $I_{\rm ^{13}CO}$ is the $^{13}$CO integrated intensity.
Then, we {converted} $\Ncoth$ to the H$_2$ column density using the abundance ratio of H$_2$ to $^{13}$CO molecules given by
\be
\NHtwo ({\rm ^{13}CO})=\Ycoth \Ncoth\ [{\rm cm^{-2}}].
\label{YN}
\ee
Here, $\Ycoth$ is adopted as $(5.0\pm 2.5)\times 10^5$ \citep{1978ApJS...37..407D} following Paper I.

%\begin{table} %%%%%%%%%%%%%%%%%%  
%\caption{Planck temperatures of the CO lines, $T_0=h \nu/k$}
%\begin{center}
%\begin{tabular}{lll} 
%\hline 
%\hline   
%Line & Freq., $\nu$ (GHz)& Planck temp., $T_0$ (K)\\
%\hline
 %\co  & 115.271204 & $T_{0}^{115}=5.53194$ \\ 
 %\coth& 110.20137 & $T_{0}^{110}=5.28864$\\
%\hline    
%\end{tabular} 
%\label{tab1} 
%\end{center}  
%\end{table} %%%%%%%%%%%%%%%%%%%

%%%%%%%%%%%%%%%% 
\begin{figure*}
\begin{center}        
\includegraphics[width=14.cm]{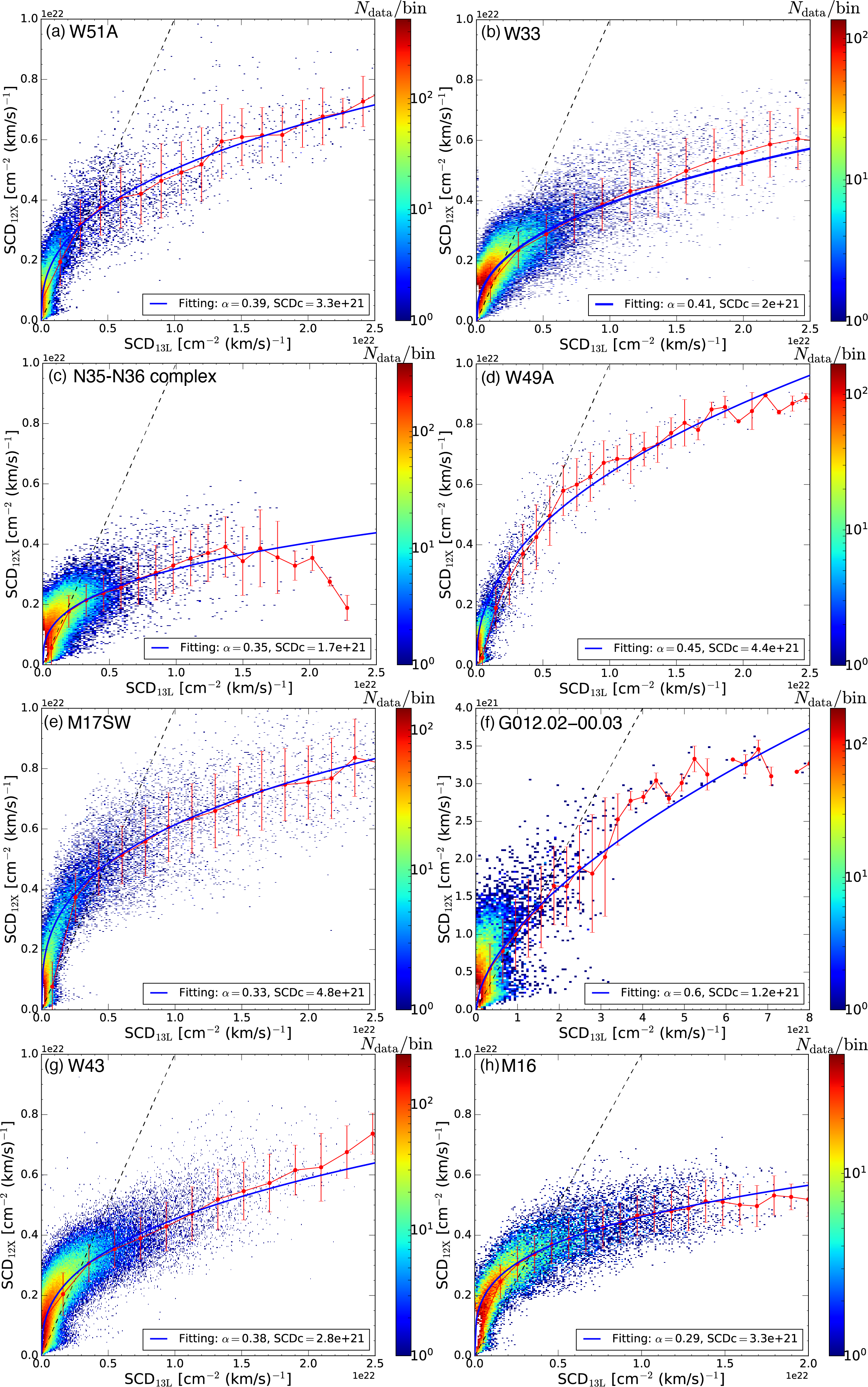}
\end{center}
\caption{Scatter {plots} between SCD$_{\rm 13L}$ and SCD$_{\rm 12X}$ of (a) W51A, (b) W33, (c) the N35-N36 complex, (d) W49A, (e) M17SW, (f) G012.02-00.03, (g) W43, and (h) M16. Red points show the averaged values of each bin, and the error bars are standard deviations of SCD$_{\rm 12X}$. Blue curves indicate the fitting results of scatter plots. The black dashed lines indicate the linear relation of SCD$_{\rm 12X}=$SCD$_{\rm 13L}$. {The color bars show the number of data points in a bin. ($N_{\rm data}/{\rm bin}$)}} 
\label{GMCscatter}
\end{figure*}
%%%%%%%%%%%%%%%%%%       

%%%%%%%%%%%%%%%% 
\begin{figure*}
\begin{center}        
\includegraphics[width=15.cm]{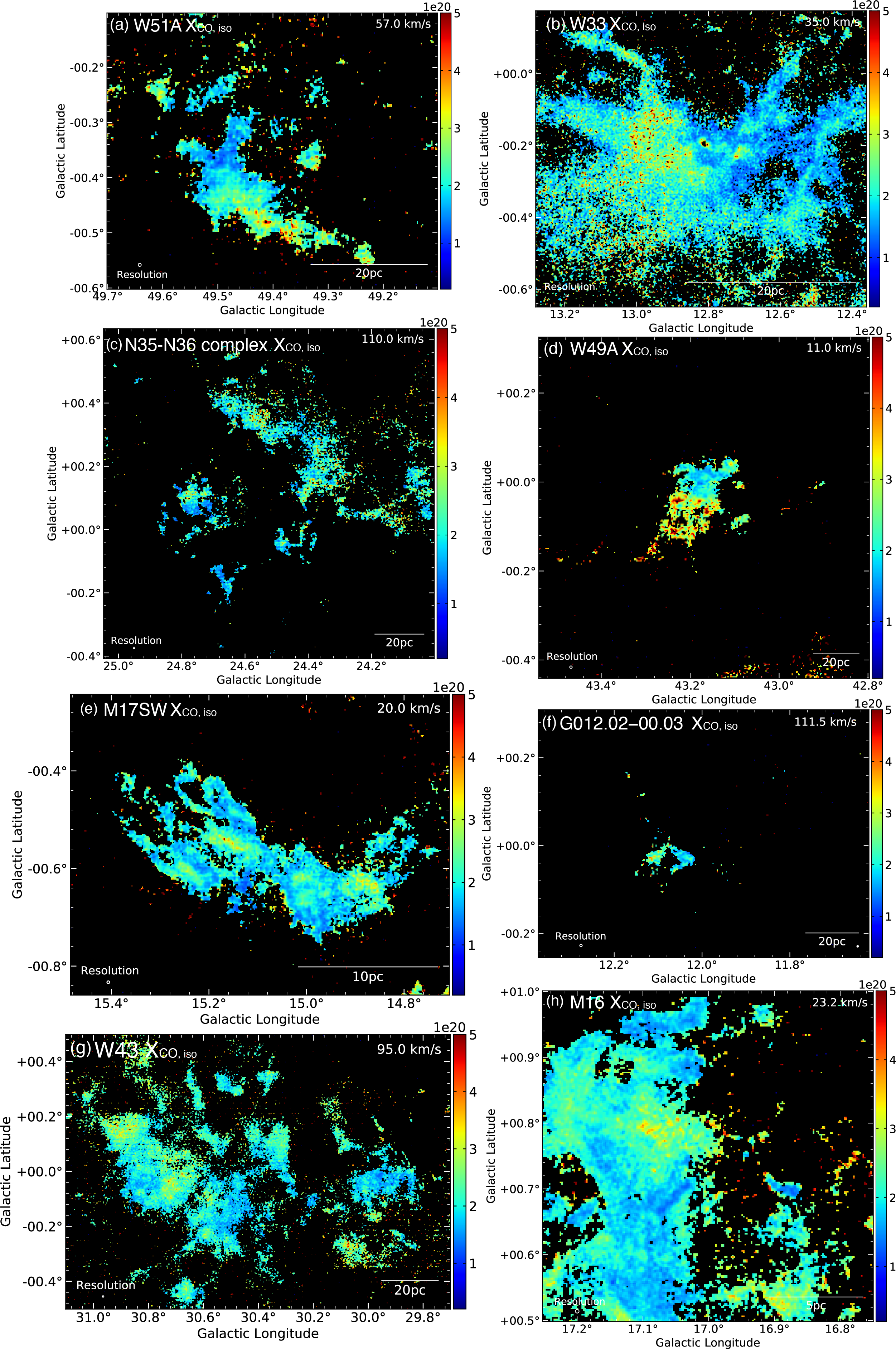}
\end{center}
\caption{
%\sof{\underline{背景を暗くすると見やすいと思います。}} 
Spatial distributions of {$X_{\rm CO, {iso}}$} in (a) W51A, (b) W33, (c) the N35-N36 complex, (d) W49A, (e) M17SW, (f) G012,02-00.03, (g) W43, and (h) M16. The points are plotted of $T_{\rm B}$ ($^{13}$CO) $> 3 \sigma$. The $1 \sigma$ noise level of each GMC is presented in Table \ref{tab1}.} 
\label{XCOmap}
\end{figure*}
%%%%%%%%%%%%%%%%%%  

%\subsection{Spectral Column Density}
We {then calculated} the H$_2$ column density per velocity channel, which is defined as the spectral column density, using {the $\Xgeneral$ and LTE method} from the intensity of $^{12}$CO and $^{13}$CO{, as described} in Paper I.
The spectral column densities (SCD) of $^{12}$CO and $^{13}$CO are expressed as
\begin{equation}
{\rm SCD}_{\rm 12X} = \frac{d \NHtwo({\rm ^{12}CO})}{dv} 
=\Xgeneral \Tbcotw\ {\rm [cm^{-2} (km\ s^{-1})^{-1}]},
\label{SCD12X}
\end{equation}
and
\begin{equation}
\begin{split}
{\rm SCD}_{\rm 13L} &= \frac{d\NHtwo({\rm ^{13}CO})}{dv} \\
&=3.0\times 10^{14}~ \frac{\tau}{1-e^{-\tau}}
\frac{\Ycoth \Tbcoth}{ 1-e^{-T_{0}^{110}/\Tex}} \ {\rm [cm^{-2} (km\ s^{-1})^{-1}]}.
\label{SCD13L}
\end{split}
\end{equation}
%Here, ${\rm SCD}_{\rm 12X}$ and ${\rm SCD}_{\rm 13L}$ are spectral column densities of H$_2$ at the peak velocity channels, derived from the $\Xgeneral$ factor and the LTE assumption, respectively.
{Here, ${\rm SCD}_{\rm 12X}$ and ${\rm SCD}_{\rm 13L}$ are the H$_2$ column densities at the peak velocity channels, assuming a standard (normalization) value of $\Xgeneral = 2.0 \times 10^{20}\ {\rm cm^{-2}\ (K\ km\ s^{-1})^{-1}}$\citep{2013ARA&A..51..207B} and $\Ycoth = 5.0\times 10^5$ \citep{1978ApJS...37..407D}, respectively. }

%%%%%%%%%%%%%%%%%%%%%%%%%%%%%%%%%%%%%%%%%%%%%%%%%
\section{Results}
%%%%%%%%%%%%%%%%%%%%%%%%%%%%%%%%%%%%%%%%%%%%%%%%%%
\subsection{{Correlation between SCD$_{\rm 12X}$ and SCD$_{\rm 13L}$ }}
Figure \ref{GMCmap}(a-h) shows the $^{12}$CO $J=$1-0 peak intensity maps of W51A, W33, N35-N36 complex, W49A, M17SW, G012.02-00.03, W43, and M16 GMC.
W49A and G012.02-00.03 GMC show compact CO distributions of $\sim 20$ pc, while W33, N35-N36 complex, W43, and M16 have more diffuse distributions of molecular gas {with about {the} half-intensity level of the peak brightness temperature.}

Figure \ref{GMCscatter}(a-h) presents scatter plots between SCD$_{\rm 12X}$ and SCD$_{\rm 13L}$ of GMCs presented in Figure\ref{GMCmap}(a-h). 
Black {dashed} lines show the linear relation of SCD$_{\rm 12X}$ = SCD$_{\rm 13L}$.
SCD$_{\rm 12X}$ shows saturation at high SCD$_{\rm 13L}$, and the nonlinear relation of each GMC has different saturation levels. {Indeed, previous studies {have} reported that the $^{12}$CO intensity shows apparent saturation at {the} high-intensity level of $^{13}$CO from the observation of Galactic molecular clouds \citep{1989ApJ...337..355L,2008ApJ...679..481P,2010PASJ...62.1277Y}. {Curve} fitting is useful {for quantitatively evaluating} the nonlinear {relationship} between the $^{12}$CO and $^{13}$CO {intensities}.}
Here, we performed curve fitting to the scatter plots between SCD$_{\rm 12X}$ and SCD$_{\rm 13L}$ using the free parameters of $\alpha$ and SCD$_{\rm c}$(spectral column density coefficient), as in Paper I.
SCD$_{\rm 12X}$ is given by
\be
\scx  = {\rm SCD}_{\rm c} \left(\frac{\scl}{{\rm SCD}_{\rm c}}\right)^\alpha.
\ee
%The fitting range was adopted of SCD$_{\rm 13L} > 3 \times 10^{21}$.
{Blue curves} show the fitting results {for} individual GMCs.
%The difference in the r.m.s noise levels in each GMC do not affect our fitting results, because there is no low-level cutoffs. 
%The data has own different physical area per pixel because each GMC is at a different distance from the solar system. The number of molecular clumps per pixel in each GMC could change drastically, which is likely to affect the opacity and ability of $\Xgeneral$ to represent the H$_2$ gas. 
{{To confirm whether the scatter plots shown in Figure \ref{GMCscatter} depend on the physical resolutions due to different distances from the GMCs,} we smoothed the data of GMCs to the same physical resolution and grid size of $\sim 1$ pc and $\sim 0.45$ pc.}
{This physical resolution corresponds to {a} beam size of $\sim 20$\arcsec at the distance of W49A, the {GMC furthest} from us in this {study}.}
{In {Figures} \ref{GMCmap_reso} and \ref{GMCscatter_reso}(a-h) of {the} Appendix, we show the obtained peak intensity maps and scatter plots of GMCs for equal physical resolution.
The results are almost the same as {those} in Figure \ref{GMCscatter}. The fitting parameters have the range of $\alpha=0.30$-$0.54$ and SCD$_{\rm c}$=$(1.6$-$4.4) \times 10^{21}{\rm cm^{-2}\ (km\ s^{-1})^{-1}}$. 
Thus, we find that the opacity and internal clump sizes in a GMC do not affect the relation of the $^{12}$CO and $^{13}$CO correlation. Our analysis also shows that the saturation level of correlations does not depend on physical spatial resolution and pixel size. }
%The correlation slopes differ from each GMC. 

%We plotted the $T_{\rm B} ({\rm ^{13}CO})> 3\sigma$ data points and the $1\sigma$ noise level of each GMC is also presented in Table \ref{tab1}.

%N35-36 complex and M17SWex show significant saturation with the diffuse distributions of molecular gas, while W51 and W49 show the large SCD$_{\rm c} \sim 3$-4 with the compact CO distribution.

%In the same system of the M17 GMC, M17SW and M17SWex have different saturation levels of SCD$_{\rm 12x}$.
%M17SWex, which is the early stage of massive star formation  (e.g., \citealp{2016ApJ...825..125P}), shows intense saturation of $^{12}$CO and $\alpha=0.12$, which is smaller than M17SW.
%These high excitation temperatures are caused by heating ultraviolet radiation from massive stars in GMCs.

%To reveal the origin of scatte differences, we performed curve fitting to these correlations.

%%%%%%%%%%%%%%%% 
\begin{figure*}
\begin{center}        
\includegraphics[width=14.cm]{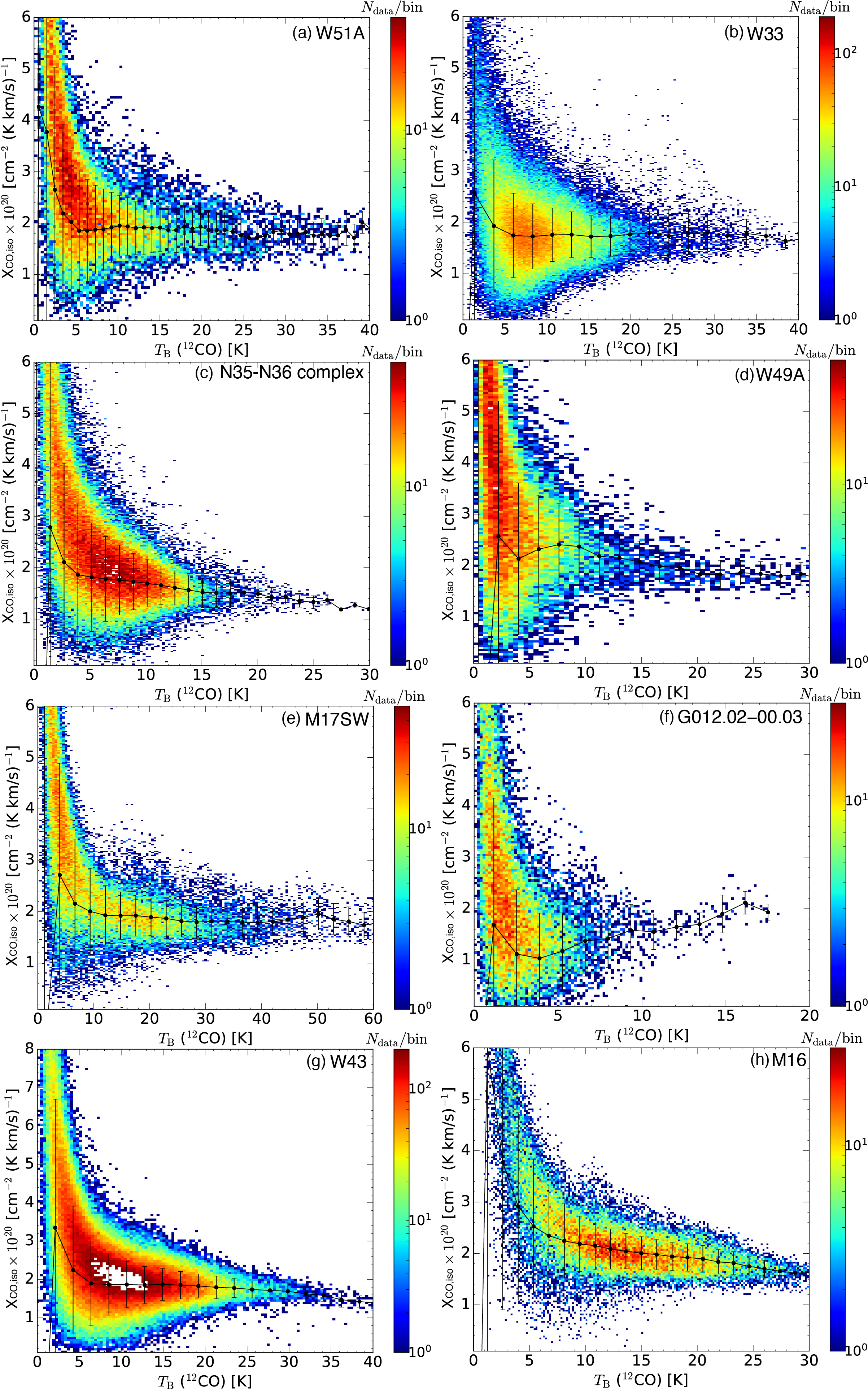}
\end{center}
\caption{Scatter {plots} between {$X_{\rm CO, {iso}}$} and $T_{\rm B}$ ($^{12}$CO) in (a) W51A, (b) W33, (c) the N35-N36 complex, (d) W49A, (e) M17SW, (f) G012,02-00.03, (g) W43, and (h) M16. Black points show the averaged values of each bin, and the error bars are standard deviations of {$X_{\rm CO, {iso}}$}. {The color bars show the number of data points in a bin. ($N_{\rm data}/{\rm bin}$)}} 
\label{XCOcor}
\end{figure*}
%%%%%%%%%%%%%%%%%%   

%%%%%%%%%%%%%%%% 
\begin{figure*}
\begin{center}        
\includegraphics[width=14.cm]{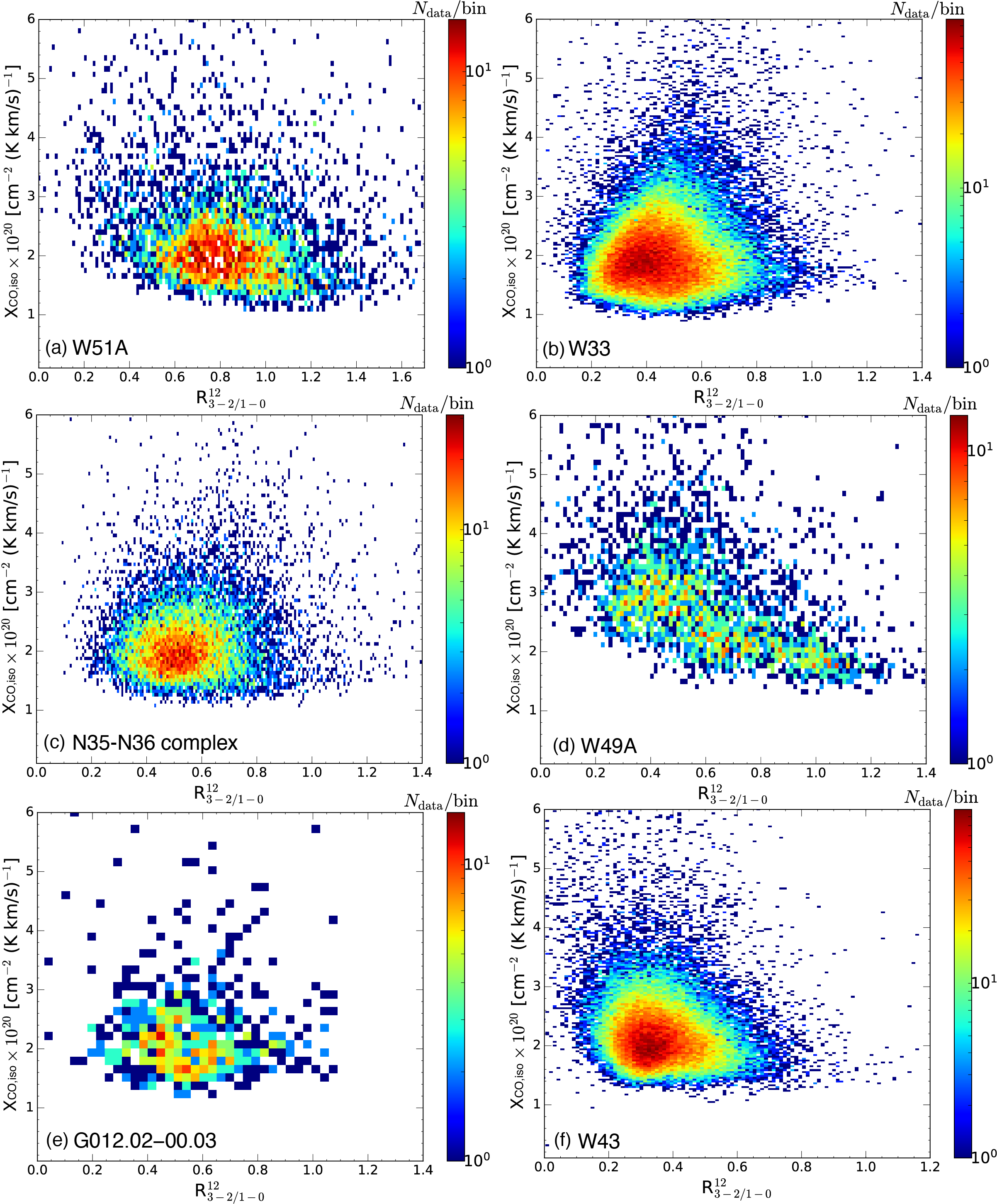}
\end{center}
\caption{{Scatter {plots} between {$X_{\rm CO, {iso}}$} and $R^{12}_{3-2/1-0}$ of (a) W51A, (b) W33, (c) the N35-N36 complex, (d) W49A, (e) G012,02-00.03, and (f) W43. The points are plotted of $T_{\rm B}$ ($^{13}$CO) $> 3 \sigma$. The color bars show the number of data points in a bin. ($N_{\rm data}/{\rm bin}$)}} 
\label{XCO_COratio}
\end{figure*}
%%%%%%%%%%%%%%%%%%   

\subsection{{Spatial distributions of {$X_{\rm CO, {iso}}$} in the Galactic GMCs}}

We obtained the $\Xgeneral$ factor {(hereafter $X_{\rm CO, {iso}}$)} of each GMC from the correlation between SCD$_{\rm 12X}$ and SCD$_{\rm 13L}$.
{$X_{\rm CO, {iso}}$} at each pixel of {the} molecular clouds is expressed as {follows:}
\be 
X_{\rm CO, {iso}} = {\rm SCD}_{\rm c} \left(\frac{\scl}{{\rm SCD}_{\rm c}}\right)^\alpha \bigg/ \Tbcotw\ {\rm [cm^{-2} (K\ km\ s^{-1})^{-1}]},
\ee
where $\alpha$ and SCD$_{\rm c}$ are the fitting parameters obtained in {Figures \ref{GMCscatter} and \ref{GMCscatter_reso}}.

Figure \ref{XCOmap}(a-h) shows the {$X_{\rm CO, {iso}}$} maps of each GMC.
{$X_{\rm CO, {iso}}$} has a variability with the range of $(0.9$-$5.0)\times 10^{20}$ [cm$^{-2}$(K km s$^{-1})^{-1}$] in a GMC.
%Comparing the $^{12}$CO peak intensity (Figure \ref{GMCmap}) with the $\Xgeneral$ map (Figure \ref{XCOmap}), we find anti-correlations between them in each GMC. 
Comparing Figure \ref{GMCmap} and Figure \ref{XCOmap}, we find an anti-correlation between the $^{12}$CO peak intensity and {$X_{\rm CO, {iso}}$}. 
Figure \ref{XCOcor} shows the scatter plot between {$X_{\rm CO, {iso}}$} and $T_B$ ($^{12}$CO) in each GMC. 
{$X_{\rm CO, {iso}}$} monotonically decreases with increasing brightness temperature, except for G012.02-00.03. It also shows a lower limit around $\Xgeneral \sim 2\times 10^{20}$cm$^{-2}$ (K km s$^{-1})^{-1}$ above $T(^{12}{\rm CO}) > 20$ K.
These results may correspond to the saturation of {the} $^{12}$CO intensity in {the} dense cores of the GMC. {This} is consistent with previous studies of Perseus molecular clouds obtained by the correlation between the $^{12}$CO integrated intensity and H$_{\rm I}$ column number density derived from the optical depth of 353 GHz dust emission ($\tau_{\rm 353}$ ) by Planck satellite observations (see Figure 12 and 15 in \citealp{Okamoto+2017}). 

\begin{table*}
%%%%%%%%%%%%%%%%%%  
\begin{center}
\caption{\revone{Fitting results and <$X_{\rm CO, {iso}}$> taken from the correlation of SCD$_{\rm 12X}$ and SCD$_{\rm 13L}$.}} 
\begin{tabular}{cccccc} 
\hline 
\hline   
Name  & $\alpha$ & SCD$_{\rm C}$ & $<X_{\rm CO, {iso}}>$ & C.C\\
&& {[$\times 10^{21}\ {\rm cm^{-2}\ (km\ s^{-1})^{-1}}$]} & {[$\times 10^{20}$ cm$^{-2}$ (K\ km\ s$^{-1})^{-1}$]}& \\
(1) & (2) & (3) & (4)  & (5)   \\
\hline
W51A  & {0.39} & {3.3} & {2.23} & {0.76}\\
W33 & {0.46} & {1.9} & {2.09} & {0.82}\\
N35-N36 & {0.34} & {1.8} & {2.20} & {0.79}\\
W49A & 0.45 & 4.4 & 2.68 & 0.78\\
M17 SW & {0.36} & {4.4} & {2.10} & {0.77}\\
G\small 012.02-00.03  & {0.54} & {1.6} & {1.69}  & {0.69}\\
W43 & {0.36} & {3.0} & {2.19}  & {0.82} \\
M16 & {0.30} & {3.3} & {2.15}  & {0.89}\\
\hline   
Average & {$0.40 \pm 0.08$} & {$2.96 \pm 1.11$} &  {$2.17 \pm 0.27$} & {$ 0.79\pm 0.06$}\\
\hline    \\
\end{tabular} 
\label{tab2}
\end{center}  
Columns: (1) GMC names (2) The fitting parameter of the non-linear relation. (3) The coefficient of the spectrum column density (SCD$_{\rm C}$). (4) The mean value of {$X_{\rm CO, {iso}}$} in a GMC. (5) Correlation coefficient. The errors of averaged value are adopted as the standard variation in all GMCs.

\end{table*} %%%%%%%%%%%%%%%%%%%

Table \ref{tab2} lists the results of parameter fitting for $\alpha$, SCD$_{\rm c}$, {$X_{\rm CO, {iso}}$}, and {the} correlation coefficient {obtained {from} the resolution of Figure \ref{GMCscatter_reso}}.
The mean value of {$X_{\rm CO, {iso}}$} of all GMCs in {this study} is {$(2.17 \pm 0.27) \times 10^{20}$} cm$^{-2}$ (K km s$^{-1})^{-1}$. {$X_{\rm CO, {iso}}$} shows the local variability in the Galactic GMCs (Figure\ref{XCOmap}), while the mean value is consistent within the {margin of error in previous works of} $2 \times 10^{20}$ cm$^{-2}$ (K km s$^{-1})^{-1}$ \citep{2013ARA&A..51..207B}, $(1.8 \pm 0.3) \times 10^{20}$\ cm$^{-2}$ (K km s$^{-1})^{-1}$ \citep{2001ApJ...547..792D}, and $(2.54 \pm 0.13)\times 10^{20}$\ cm$^{-2}$ (K km s$^{-1})^{-1}$ \citep{2011A&A...536A..19P}.

%%%%%%%%%%%%%%%%%%%%%%%%%%%%%%%%%%%%%%%%%%%%%%%%%
\section{{Discussion}}
%%%%%%%%%%%%%%%%%%%%%%%%%%%%%%%%%%%%%%%%%%%%%%%%%%

\subsection{{The variability of {$X_{\rm CO, {iso}}$} }}
{Our fitting results show that the correlation of column densities between $^{12}$CO and $^{13}$CO {is} not universal in Galactic GMCs. Each GMC has its own correlation parameters ({Figures} \ref{GMCscatter} and \ref{GMCscatter_reso}). \citet{2001A&A...365..571W} reported that the $\Xgeneral$ factor for a virialized GMC depends on the kinetic temperature ($T_{\rm kin}$) and molecular hydrogen number density ($n({\rm H_2})$) of {the} molecular gas. (see also Chapter 2.1 in \citealp{2013ARA&A..51..207B}). 
{To reveal} the origin of the variations {in} {$X_{\rm CO, {iso}}$}, we investigated the $^{12}$CO $J=$3-2/1-0 intensity ratio (hereafter $R^{12}_{3-2/1-0}$) using the JCMT $^{12}$CO $J=$3-2 archival data \citep{2013ApJS..209....8D,2023ApJS..264...16P}. $R^{12}_{3-2/1-0}$ depends on $T_{\rm kin}$ and $n({\rm H_2})$ in molecular clouds assuming the large velocity gradient model \citep{1974ApJ...189..441G}. 
We {analyzed} the W51, W33, N35, W49A, G12.02-00.03, and W43 GMC because the JCMT data {are} only covered with $|b|<0.5 \deg$.
Figure \ref{XCO_COratio} shows the scatter plot between {$X_{\rm CO, {iso}}$} and $R^{12}_{3-2/1-0}$. 
The plots of each GMC do not show a clear correlation. Thus, we suggest that the origin of own correlation parameters and variations of {$X_{\rm CO, {iso}}$} might not be simply density and/or temperature variations in  a molecular cloud.
{In our analysis, we assumed that $\Ycoth$ is constant in all GMCs. We point out that the change of the isotope abundance ratio with density might explain the variability of $X_{\rm CO, {iso}}$ in a GMC.}
%Stellar feedback from HII region and massive stars embedded in a molecular cloud can affect the physical conditions of molecular gas. The kinetic temperature and hydrogen number density change with $\lesssim 10$ pc scale in a GMC holding star-forming activity (e.g., Figure 20 in \citealp{2015ApJS..216...18N}). We point out that the difference of local stellar feedback with $\lesssim 10$ pc scale in each GMC might explain the origin of the variability of the $\Xgeneral$ factor.}

%\subsection{Correlation between the $\Xgeneral$ factor and $^{12}$CO brightness temperature}
%\subsection{Dependence of the isotope abondance ratio}
%CO isotope aboundance ratio varried in a molecular cloud because of the  UV radiation from the OB-type stars. 

\subsection{Radial gradient of the $\Xgeneral$ factor}
%%%%%%%%%%%%%%%% 
\begin{figure}
\begin{center}        
\includegraphics[width=8.cm]{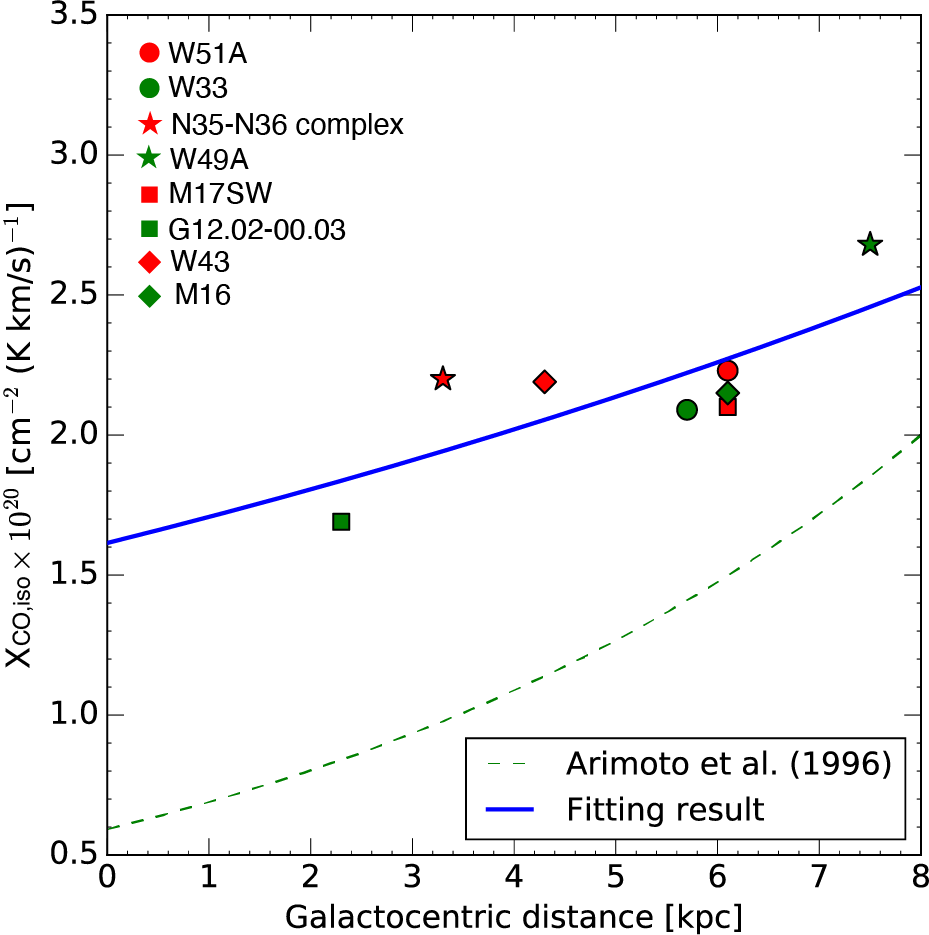}
\end{center}
\caption{The {$X_{\rm CO, {iso}}$} gradient from the distance of the Galactic Center. The blue line shows the fitting result adopted by the exponential function. The green dashed line indicates the relation from \citet{arimoto+1996}.}
\label{XCO_cor}
\end{figure}
%%%%%%%%%%%%%%%%%%   
{In addition}, we investigated the radial gradient of the $\Xgeneral$ factor in the Galactic disc. The averaged $\Xgeneral$ factor has a high value in W49A at Galactocentric distance $R=$7.5 kpc, while G012.02-00.03 has a low value $R=$2.3 kpc.
{According to \cite{arimoto+1996}, the $\Xgeneral$ factor increases with the Galactocentric radius by an exponential function given by
\begin{eqnarray}
{\rm log_{10}}\left( \frac{\Xgeneral}{X_0}\right) &=& 0.41 \frac{(R-R_0)}{r_e},
\end{eqnarray}
where $r_e=6.2$ kpc is the scale radius of {the} Galactic disc and $X_0$ is $\Xgeneral$ at $R=R_0 = 8.0$ kpc \citep{2020PASJ...72...50V,2019ApJ...885..131R}. 
This equation can be deformed as follows:
\begin{eqnarray}
\Xgeneral %&=&{2.0\times 10^{20} \exp \left({R-R_0 \over 6.567\ {\rm [kpc]}}\right)}\\
={0.59 \times 10^{20} \exp \left(\frac{R}{6.567\ {\rm [kpc]}}\right)}\ {\rm [cm^{-2} (K\ km\ s^{-1})^{-1}]}
\end{eqnarray}
}

Figure \ref{XCO_cor} shows a plot of {$X_{\rm CO, {iso}}$} obtained in this paper as a function of the Galactocentric distance. 
It is found that {$X_{\rm CO, {iso}}$} increases with the distance from the Galactic Centre, and the plot may be fitted by an exponential function by 
\be 
{X_{\rm CO, {iso}} = 1.61\times 10^{20}\exp\left({R \over 17.85 [{\rm kpc}]}\right){\rm [cm^{-2} (K\ km\ s^{-1})^{-1}]}}.
\ee
%The green dotted line presents the radial gradient obtained by \cite{arimoto+1996}. 
%Our fitting result shows the translation along the y-axis ($\Xgeneral$) positive direction about a factor 3 compared with \cite{arimoto+1996}. 
{Previous studies reported that a value of the $\Xgeneral$ factor depends on the metal abundance in galaxies (e.g., \citealp{arimoto+1996,1997A&A...328..471I,Leroy+2011,2012ApJ...746...69G}).
We suggest that the galactocentric dependency of {$X_{\rm CO, {iso}}$} is caused by the radial metallicity gradient from the Galactic center to {the} outer Galaxy in the Milky Way. }
Our fitting result yields a significantly larger $\Xgeneral$ at the Galactic Centre by a factor of 3 compared to that by \cite{arimoto+1996} indicated by the green {dashed} line.
This discrepancy may be caused by no data point within $R < 2$ kpc in our study
as well as by the normalization of the local value to $2\times 10^{20}$ \uxco at 8 kpc in the previous study.
\cite{1995ApJ...452..262S} reported that the $\Xgeneral$ factor within 400 pc in the Galactic Centre region is by a factor 3-10 lower than the Galactic disk.
Therefore, the value of $\Xgeneral$ in the Galactic Centre is still controversial. 
The present method applied to the $^{12}$CO and $^{13}$CO $J =$1-0 line data in the Galactic Centre region would provide a more precise determination of $\Xgeneral$ in the innermost Milky Way.
A detailed analysis of this point will be presented in a separate paper using the CO survey data of the Galactic Centre (e.g., \citealp{1998ApJS..118..455O,2010PASJ...62.1307T,2014ApJ...780...72E,2019PASJ...71S..19T}).  

\subsection{{Comparison with the virial methods}}

\begin{table*}
{
%%%%%%%%%%%%%%%%%%  
\begin{center}
\caption{{Physical parameters and $X_{\rm CO, {vir}}$ using virial method}} 
\begin{tabular}{ccccccccc} 
\hline 
\hline   
Name & $R$ & $\sigma_v$  & $M_{\rm vir}$ & $L_{\rm CO}$ &$X_{\rm CO, {vir}}$ &{$X_{\rm CO,iso}/X_{\rm CO,vir}$}\\
& [pc] & [\kms] &  [$M_{\odot}$] & [K km s$^{-1}$ pc$^2$] & {[$\times 10^{20}$ cm$^{-2}$ (K\ km\ s$^{-1})^{-1}$]} & \\
(1) & (2) & (3) & (4)  & (5) & (6) &(7)\\
\hline
W51A & 6.3 & 4.2 & $1.2 \times 10^5$ & $3.5 \times 10^4$& 1.51 &1.5\\
W33 & 6.5 & 2.5 & $4.3 \times 10^4$ & $1.7 \times 10^4$ & 1.15 &1.8\\
N35-N36 & 32 & 4.4 & $6.2 \times 10^5$ &$3.8 \times 10^5$& 0.73 &3.0\\
W49A & 4.8 & 4.8 & $1.2 \times 10^5$ & $2.3 \times 10^4$ & 2.21 &1.2\\
M17 SW & 2.6 & 3.0 & $2.4 \times 10^4$ & $6.1 \times 10^3$ & 1.74 &1.2\\
G\small 012.02-00.03 & 7.1 & 2.4 & $4.2 \times 10^4$ & $7.7 \times 10^3$ & 2.47 &0.68\\
W43 & 20 & 4.0 & $3.2 \times 10^5$ & $2.5 \times 10^5$ & 0.58 &3.8 \\
M16 & 6.0 & 2.1 & $2.9 \times 10^4$ & $1.3 \times 10^4$ & 1.01 &2.1\\
\hline   
Average & $11$ & $3.4$ &  $1.7\times 10^5 $ & $9.2 \times 10^4$& $1.43\pm 0.68$ &$1.80\pm 0.52^*$\\
\hline\\
\end{tabular} 
\label{tab3}
\end{center}
Columns: (1) GMC names (2) Cloud radius defined by the \%30 levels of the peak integrated intensity (3) Mean velocity dispersion in a molecular cloud (4) Virial mass (5) $^{12}$CO luminosity (6) The $\Xgeneral$ factor derived by the virial mass. {(7) Ratio of $X_{\rm CO, iso}$ to $X_{\rm CO, vir}$. *The averaged value is calculated from the ratio of averaged $X_{\rm CO, iso}$ and $X_{\rm CO, vir}$.} }
\end{table*} %%%%%%%%%%%%%%%%%%%

%\begin{figure}
%\begin{center}         
%\includegraphics[width=8cm]{XisoXvirRatio.pdf}
%\includegraphics[width=8cm]{XCOratiovsCOluminosity_ver2.png}
%\includegraphics[width=8cm]{XcoXvRatio.jpg}
%\end{center}
%\caption{\sof{$X_{\rm CO:vir}$ (circles), $X_{\rm CO:iso}$ (diamonds), and ratio of $X_{\rm CO, iso}$ to $X_{\rm CO, vir}$ (triangles) as a funciton of the CO luminosity $L_{\rm CO}$. The blue line shows a fit to the plot.} }
%\label{XisoXvirRatio}
%\end{figure}

{Finally, we calculated the $\Xgeneral$ factor from the virial mass of each GMC compared with our estimation. According to \citet{1987ApJ...319..730S}, the virial mass of GMCs is given by
\begin{equation}
M_{\rm vir} = 1040\ R\ \sigma_v^2\ [M_{\odot}],
\label{virial}
\end{equation}
where $R$ is {the} effective {radius} of a GMC in {parsec} and $\sigma_v$ is the 1D velocity dispersion in \kms, assuming a density profile of $\rho (r) \propto r^{-1}$. 
We estimated the radius of a GMC with $R=\sqrt{S/\pi}$, where S is the cloud area within 30\% levels of the $^{12}$CO peak integrated intensity. The velocity dispersion is adopted as the averaged value of each pixel in a GMC.}
{Then, the $\Xgeneral$ factor using virial methods is expressed as
\begin{equation}
X_{\rm CO, vir} = {M_{\rm vir} \over \mu_{\rm H_2} m_{\rm H}L_{\rm CO}}\ {\rm [cm^{-2} (K\ km\ s^{-1})^{-1}]},
\label{virial_XCO}
\end{equation}
where $\mu_{\rm H_2} \sim 2.8$ is the mean molecular weight per hydrogen molecule (Appendix A.1 in \citealp{2008A&A...487..993K}), $m_{\rm H}=1.67 \times 10^{-24}$g is proton mass and $L_{\rm CO}$ is the $^{12}$CO total luminosity. Table \ref{tab3} shows the results of the $\Xgeneral$ factor in each GMC obtained by the virial methods. The ratio of averaged $\Xgeneral$ taken from the virial mass and CO isotopologues is $1.80\pm0.52$. }
\section{Summary}
%%%%%%%%%%%%%%%%%%%%%%%%%%%%%%%%%%%%%%%%%%%%%%%%%%%%%%
The conclusions of this paper are summarized as follows:
\begin{enumerate}
\item We studied the correlation between $^{12}$CO and $^{13}$CO intensities toward the Galactic GMCs W51A, W33, N35-N36 complex, W49A,  M17SW, G12.02-00.03, W43, and M16 using the FUGIN CO survey data taken with the Nobeyama 45 m telescope. 
\item All the GMCs show intensity saturation of the $^{12}$CO line in regions with high brightness of $^{13}$CO. 
\item We also {present} high-resolution {$X_{\rm CO, {iso}}$} maps made from the correlations of the CO isotopologues, which revealed local variability of {$X_{\rm CO, {iso}}$} in each GMC. We also {show} that the {$X_{\rm CO, {iso}}$} monotonically decreases with increasing $^{12}$CO brightness temperature.
\item The averaged value of all GMCs is calculated to be $X_{\rm CO, {iso}} = {(2.17 \pm 0.27)\times 10^{20}}$ cm$^{-2}$ (K\ km\ s$^{-1})^{-1}$, which is consistent {within the margin of} error with the reported values in the previous review.
\item We {show} that {$X_{\rm CO, {iso}}$} increases with the Galactocentric distance in accordance with the previous works while suggesting a {larger} value {for} the {Galactic Centre} by a factor of 3 {compared to that by \cite{arimoto+1996}}. 
{\item The averaged value of the $\Xgeneral$ factor taken from the virial mass and CO isotopologues of GMCs is consistent within the error by a factor of 2.}
%Our fitting result shows the translation along the y-axis ($\Xgeneral$) about a factor 3 comparing with \cite{arimoto+1996}
\end{enumerate}

\section*{Acknowledgements}
{The authors are grateful to the anonymous referee for carefully reading our manuscript and giving us thoughtful suggestions, which greatly improved this paper. We are grateful to Dr. Shinji Fujita of the University of Tokyo and Prof. Toshihiro Handa of Kagoshima University for useful comments on data analysis and discussion. The authors would like to thank the FUGIN project members for the CO data observed by the Nobeyama 45-m telescope.}
The Nobeyama 45-m radio telescope is operated by the Nobeyama Radio Observatory.
We utilized the Python software package for astronomy \citep{2013A&A...558A..33A}, NumPy \citep{2011CSE....13b..22V}, Matplotlib \citep{2007CSE.....9...90H}, IPython \citep{2007CSE.....9c..21P}, CASA \citep{2022PASP..134k4501C}, and APLpy \citep{2012ascl.soft08017R}. 
%The authors are grateful to the anonymous referee for the valuable comments.

\vskip 2mm

\noindent{\bf Data availability:} The data underlying this article are available in \href{http://nro-fugin.github.io}{http://nro-fugin.github.io}. {The FUGIN CO data were retrieved from the JVO portal} \href{http://jvo.nao.ac.jp/portal}{http://jvo.nao.ac.jp/portal} 
{operated by ADC/NAOJ.}
{The JCMT COHRS cube data is downloaded from the CANFAR \href{https://www.canfar.net/storage/vault/list/AstroDataCitationDOI/CISTI.CANFAR/22.0078/data/RELEASE2_CUBE_REBIN}{data archive}\citep{2009ASPC..411..185G}.}

\vskip 2mm
\noindent{\bf Conflict of interest:} The authors declare that there is no conflict of interest.

\begin{appendix}

\section{{Effect of the physical resolution }}
%%%%%%%%%%%%%%%% 
%\sof{\underline{Gridではなくbeam (resolution))依存はどうかという}} \\
%\sof{\underline{意味ではないでしょうか。}}\\
%\sof{\underline{ナイキストサンプリングよりこまかいグリッドを}}\\
%\sof{\underline{どう変えても何もかわらないと思います。}}\\

{{To} confirm that the scatter plots shown in Figure \ref{GMCscatter} do not depend on the physical {resolutions} due to different distances to the GMCs, we smoothed the data and made the same physical resolution maps of $\sim 1$ pc (Figure \ref{GMCmap_reso}).
The result is shown in Figure \ref{GMCscatter_reso}, which {is} almost identical to the original plots. We thus confirm that the results are not affected by the different distances of the objects.}

\begin{figure*}
\begin{center}        
\includegraphics[width=15cm]{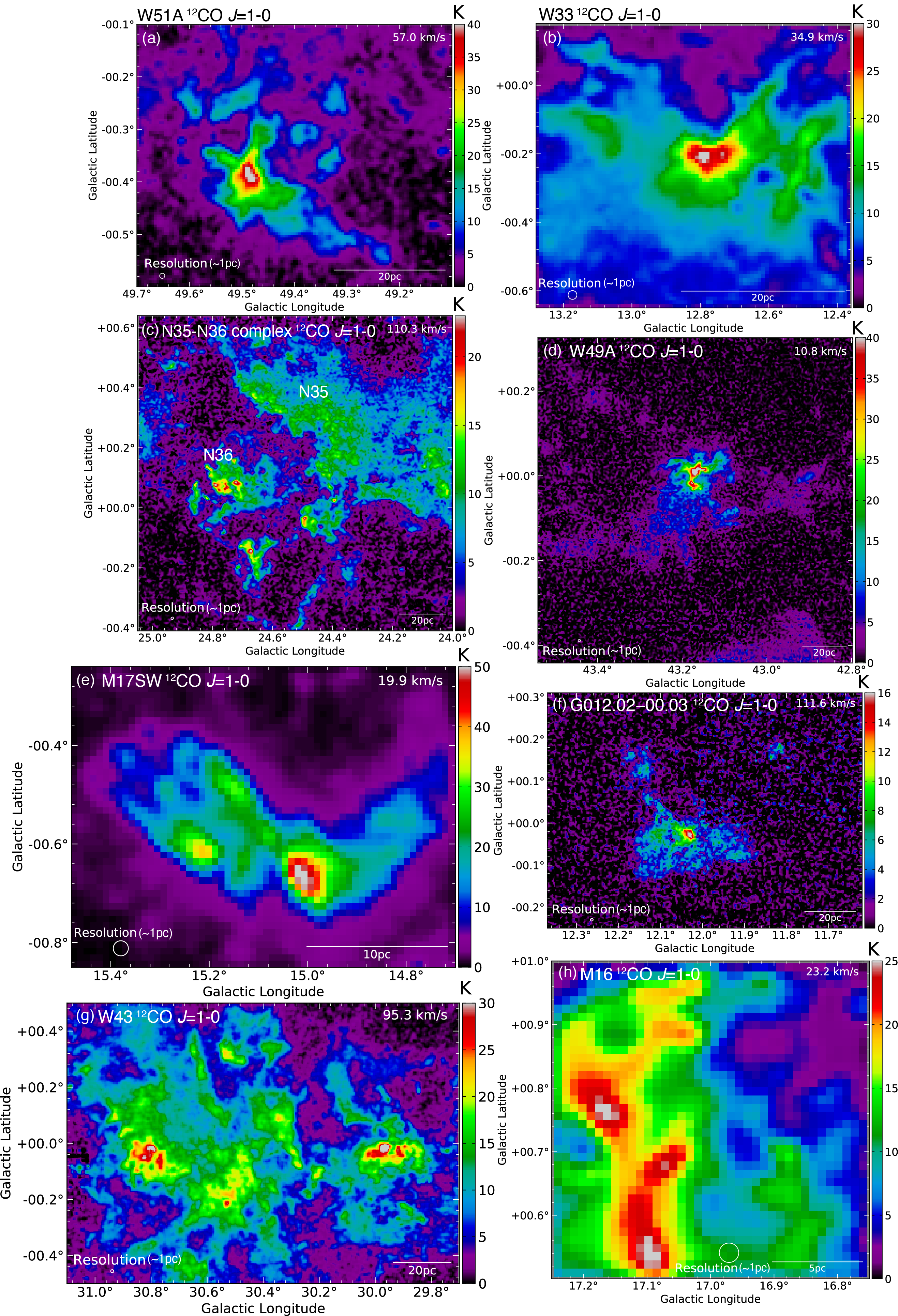}
\end{center}
\caption{{Same as Figure \ref{GMCmap}, but the data was smoothed to the same physical resolution of $\sim 1$ pc.} } 
\label{GMCmap_reso}
\end{figure*}

\begin{figure*}
\begin{center}        
\includegraphics[width=15.cm]{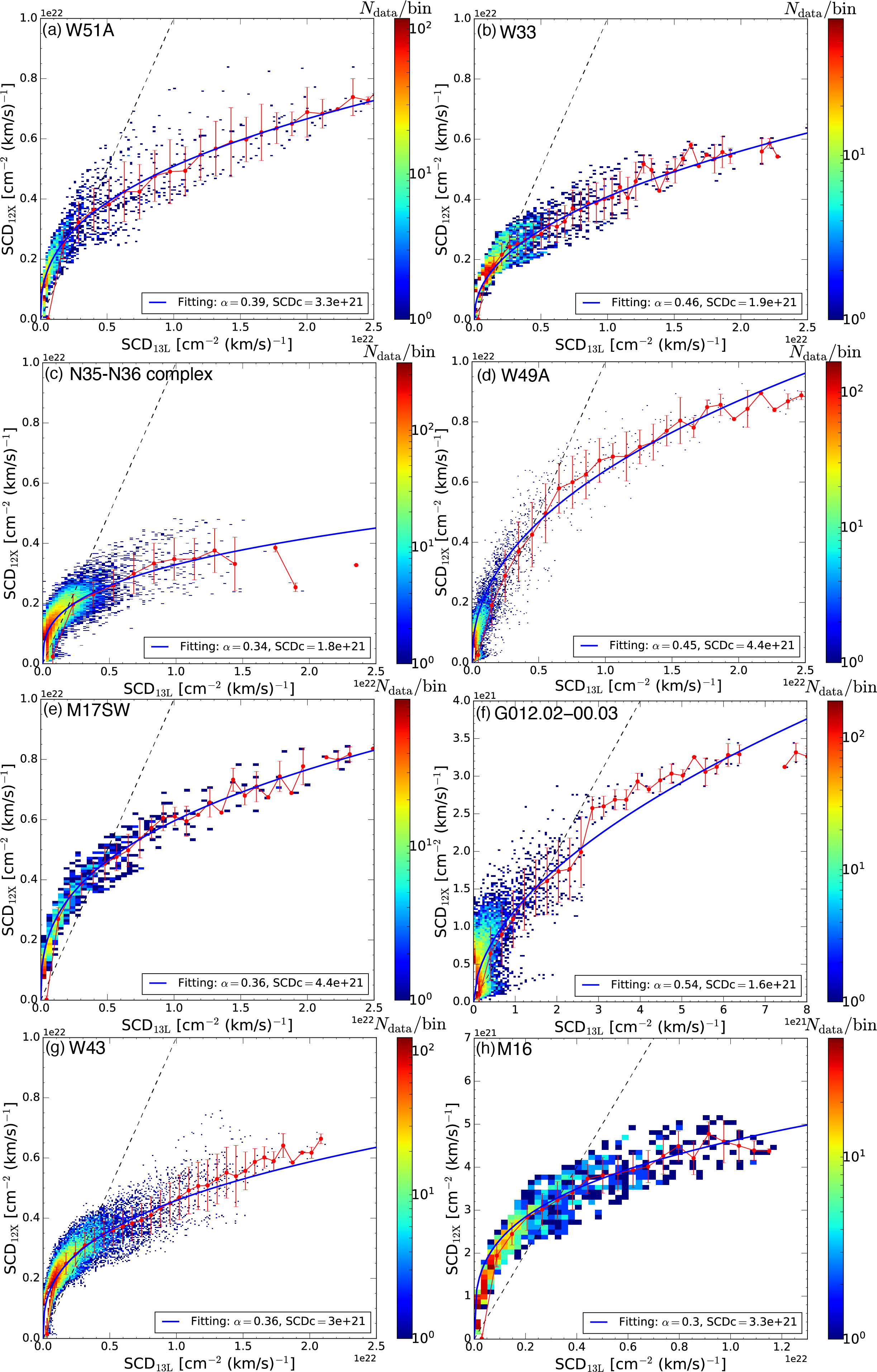}
\end{center}
\caption{{Same as Figure \ref{GMCscatter}, but the data was smoothed to the same physical resolution of $\sim 1$ pc.} } 
\label{GMCscatter_reso}
\end{figure*}
%%%%%%%%%%%%%%%%%%   
\end{appendix}
\end{document}